\documentclass[sigconf]{acmart}

\usepackage{amsmath}
\usepackage{todonotes}
\usepackage{booktabs}
\usepackage{subfig}

\DeclareMathOperator*{\argmax}{arg\,max}

\AtBeginDocument{
  }

\copyrightyear{2026}
\acmYear{2026}
\setcopyright{cc}
\setcctype{by}
\acmConference[WWW '26]{Proceedings of the ACM Web Conference 2026}{April 13--17, 2026}{Dubai, United Arab Emirates}
\acmBooktitle{Proceedings of the ACM Web Conference 2026 (WWW '26), April 13--17, 2026, Dubai, United Arab Emirates}
\acmPrice{}
\acmDOI{10.1145/3774904.3792914}
\acmISBN{979-8-4007-2307-0/2026/04}

\begin{document}

\title{Efficient Learning of Sparse Representations from Interactions}

\author{Vojtěch Vančura}
\email{vojtech.vancura@recombee.com}
\orcid{0000-0003-2638-9969}
\affiliation{
  \institution{Recombee}
  \city{Prague}
  \country{Czech Republic}
}
\additionalaffiliation{
  \institution{Faculty of Mathematics and Physics, Charles University}
  \city{Prague}
  \country{Czech Republic}
}

\author{Martin Spišák}
\email{martin@topk.io}
\orcid{0009-0006-7763-5575}
\affiliation{
  \institution{TopK}
  \city{San Francisco}
  \state{CA}
  \country{USA}
}
\authornote{Work completed while at Recombee.}
\authornotemark[1]

\author{Rodrigo Alves}
\email{rodrigo.alves@fit.cvut.cz}
\orcid{0000-0001-7458-5281}
\affiliation{
  \institution{Czech Technical University in Prague}
  \city{Prague}
  \country{Czech Republic}
}

\author{Ladislav Peška}
\email{ladislav.peska@matfyz.cuni.cz}
\orcid{0000-0001-8082-4509}
\affiliation{
  \institution{Faculty of Mathematics and Physics, Charles University}
  \city{Prague}
  \country{Czech Republic}
}

\begin{abstract}
Behavioral patterns captured in embeddings learned from interaction data are pivotal 
across various stages of production recommender systems. However, in the initial retrieval stage, practitioners face an inherent tradeoff between embedding expressiveness and the scalability and latency of serving components, resulting in the need for representations that are both compact and expressive. 

To address this challenge, we 
propose a training strategy for learning high-dimensional sparse embedding layers in place of conventional dense ones, balancing efficiency, representational expressiveness, and interpretability. 
To demonstrate our approach, we modified the production-grade collaborative filtering autoencoder ELSA, achieving up to $10\times$ reduction in embedding size with no loss of recommendation accuracy, and up to $100\times$ reduction with only a 2.5\% loss. Moreover, the active embedding dimensions reveal an interpretable inverted-index structure that segments items in a way directly aligned with the model’s latent space, thereby enabling integration of segment-level recommendation functionality (e.g., 2D homepage layouts) within the candidate retrieval model itself. Source codes, additional results, as well as a live demo are available at \url{https://github.com/zombak79/compressed\_elsa}.

\end{abstract}



\maketitle

\section{Introduction}
From pure collaborative filtering~\cite{
5197422, 
10.1145/3178876.3186150} 
to sophisticated sequential approaches~\cite{
8594844, 10.1145/3357384.3357895}
, a large body of recommender systems (RSs) research aims to design effective algorithms for learning representations from user--item interactions. This sustained interest stems from the fact that such representations play a central role in production-scale RSs. First, they are crucial in early-stage retrieval, where their quality directly determines which items are surfaced for subsequent ranking stages. Second, these representations are often reused across downstream tasks (e.g., user or item clustering), making their structure and expressiveness broadly impactful. However, in large-scale retrieval scenarios, these representations introduce significant practical challenges. Modern RSs must handle massive item catalogs, yet embedding dimensionality is tightly constrained by serving latency, memory footprint, and compute budgets. As a result, practitioners face a persistent tension: while larger and more expressive embeddings typically improve accuracy, production constraints necessitate compact representations, motivating techniques that balance expressiveness, efficiency, and scalability.

\paragraph{\textbf{Related Work}}
Recent advances in document retrieval highlight \emph{sparse representations} as a powerful alternative to dense embeddings. A prominent example is SPLADE~\cite{10.1145/3404835.3463098}, which learns sparse representations of text as lexical expansions aligned with vocabulary terms, enabling inverted–index retrieval while achieving accuracy comparable to state-of-the-art dense methods~\cite{lassance2024spladev3newbaselinessplade}.

A complementary line of work explores \emph{sparse high-dimensional projections} of dense embeddings~\cite{wen2025matryoshkarevisitingsparsecoding,compressae}. By constraining only the number of active dimensions, these approaches achieve strong compression with minimal quality loss and have been shown to outperform Matryoshka-style low-dimensional embeddings~\cite{10.5555/3600270.3602462} of equal size. The advantage is supported by theoretical results showing that sparse codes in large latent spaces can represent exponentially more distinct patterns than dense vectors at comparable storage budget~\cite{baldi2021theorycapacitysparseneural}, which can be viewed as providing greater effective dynamic range. Moreover, the resulting activation patterns naturally support efficient inverted-index retrieval algorithms~\cite{10.1145/3626772.3657769}.

RSs face a challenge distinct from document retrieval: rather than operating over a fixed vocabulary of tokens, they maintain embeddings for millions of users and items that evolve over time. Unsurprisingly, these embedding tables dominate model size in modern deep recommenders~\cite{10.1145/3637841}, resulting in a memory bottleneck for downstream model inference. As a result, RSs stand to benefit substantially from compact yet expressive representations.

\paragraph{\textbf{Contributions}}
We introduce an approach that learns \emph{sparse entity embeddings} during collaborative filtering model training, rather than via post-hoc compression.
As a backbone method, we adopt ELSA~\cite{elsa}, a linear autoencoder widely deployed in industry 
 due to its state-of-the-art retrieval performance and scalability~\cite{sparse_elsa}. 
ELSA is particularly suitable for this study because (1) architecturally, it consists of the core components shared by many neural RS architectures -- item embedding and de-embedding layers -- while replacing the intermediate network with a simple pooling operation over the embeddings of items in the interaction sequence; and (2) as a shallow model, it must encode all representational power in its embeddings. As a result, ELSA typically relies on relatively large embedding sizes
~\cite{elsa,sparse_elsa}, leaving substantial room for compression.

Our experiments validate the advantages suggested by prior work on sparse representations: they achieve a highly compelling quality--size tradeoff. 
Specifically, our \textit{Compressed ELSA} model attains accuracy comparable to standard (dense) ELSA despite orders-of-magnitude smaller embedding size, and outperforms low-dimensional ELSA, post-hoc sparse ELSA compression via CompresSAE~\cite{compressae}, and even the highly efficient pruned EASE model~\cite{ease} at equal storage budgets.
Finally, the learned sparse embeddings exhibit interpretability, and their inverted-index structure  reveals coherent item segments. These segments can be used for candidate prefiltering and for constructing segment embeddings aligned with the model’s latent space, enabling unified recommendation over items and segments, as we demonstrate in our online demo.

\section{Method}\label{method}

\textbf{\textit{ELSA}}.
Let $\boldsymbol X \in \{0,1\}^{m \times n}$ be the training interaction matrix between $m$ users and $n$ items, where $\boldsymbol X_{i,j}=1$ indicates that user $i$ interacted with item $j$, and 0 otherwise.
In this setting, \textbf{ELSA}~\cite{elsa, sparse_elsa} optimizes 

\begin{equation}
    \label{eq:loss_elsa_gen}
    \text{min}_{\boldsymbol{A}} \; \mathcal{L} \!\left( \boldsymbol{X}, \, \!\boldsymbol{X}\left(\boldsymbol{A}\boldsymbol{A}^{\top} - \boldsymbol{I} \right)\right),
\end{equation}

\noindent where $\boldsymbol{A} \in \mathbb{R}^{n \times d}$ is a (\emph{dense}) item embedding matrix with rows constrained to unit $\ell_2$-norm, and $\boldsymbol{I} \in \mathbb{R}^{n \times n}$ is the identity matrix. 
Given a user interaction vector $\mathbf{x} \in \{0,1\}^n$, ELSA predicts the relevance scores $\hat{\mathbf{r}}$ as 
$
        \hat{\mathbf{r}}^\top = \mathbf{x}^{\top} \boldsymbol{A} \boldsymbol{A}^{\top} - \mathbf{x}^{\top}
$.
For further details (including the definition of the loss function $\mathcal{L}$), we refer the reader to~\cite{sparse_elsa}.

\paragraph{\textbf{Compressed ELSA}} To learn \emph{sparse} item embeddings, we modify the original ELSA formulation in~\eqref{eq:loss_elsa_gen} by enforcing row-wise sparsity on $\boldsymbol{A}$. For $k \in \mathbb{N}$, let $\mathcal{S}_k(\boldsymbol{A}) = \text{mask}(\boldsymbol{A}, k) \odot \boldsymbol{A}$ denote a row-wise deterministic (absolute) top-$k$ sparsification operator defined via 

    $$
        \text{mask}(\boldsymbol{A}, k)[i, j] = \begin{cases}
            1 & \text{if }\bigl|\boldsymbol{A}\bigr|[i,j]\text{ is a top-}k\text{ element in its row }\\
            0 & \text{otherwise}.
        \end{cases}
    $$

Applying $\mathcal{S}_k$ to a dense matrix $\boldsymbol{A}$ yields a sparse matrix
$\boldsymbol{A}_{s}$, for which we re-normalize each row to unit $\ell_2$-norm, obtaining $\bar{\boldsymbol{A}}_s$. Using this notation, we formulate the Compressed ELSA's optimization and inference as follows, respectively: 
\begin{align}
&\text{min}_{\boldsymbol{A}} \; \mathcal{L} \!\left( \boldsymbol{X}, \, \!\boldsymbol{X}\left(\bar{\boldsymbol{A}}_s\bar{\boldsymbol{A}}_s^{\top} - \boldsymbol{I} \right)\right) \text{ and }
\hat{\mathbf{r}}^\top = \mathbf{x}^{\top} \bar{\boldsymbol{A}}_s \bar{\boldsymbol{A}}_s^{\top} - \mathbf{x}^{\top}.
\end{align}

\begin{figure*}[tb]
    \centering
    \includegraphics[width=\textwidth]{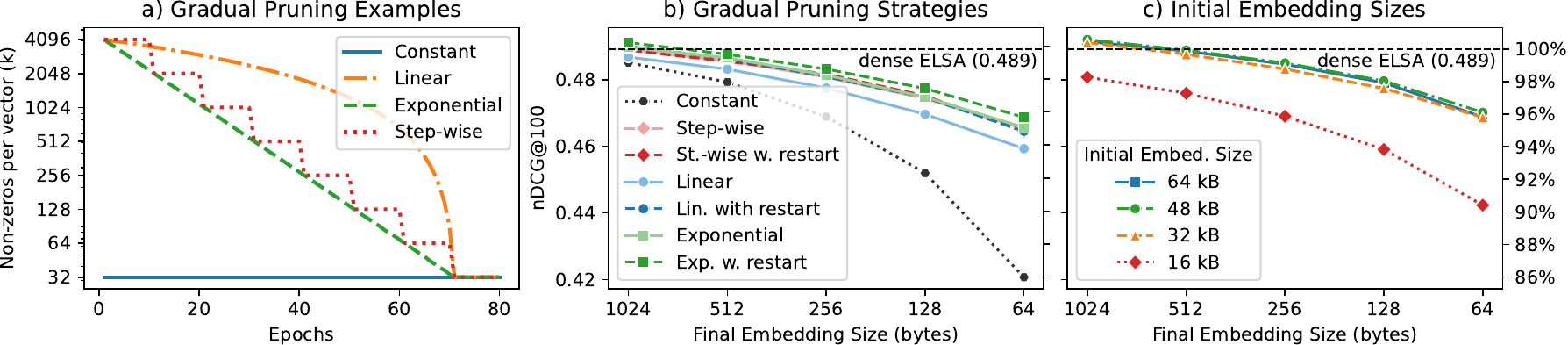}
    \caption{Ablation study: {\normalfont{(a) Comparison of gradual pruning strategies; (b) Performance under different pruning strategies with and without training restarts; (c) Effect of initial embedding dimensionality. All results report nDCG@100 on the \textit{Goodbooks-10k} dataset.}}}
    \label{fig:results}
\end{figure*}

\paragraph{\textbf{Pruning strategies}} A direct application of the sparsification operator with a fixed target
$k \in \mathbb{N}$ from the start is likely to cause a portion of latent dimensions to stop activating during training -- an effect analogous to the \emph{dead-latents} phenomenon~\cite{gao2024scalingevaluatingsparseautoencoders} observed in sparse autoencoders. To avoid this
degeneration, we introduce pruning schedules that gradually decrease the number
of allowed nonzero entries. Formally, we define a sequence of sparsity
levels $\{k_t\}_{t=0}^{T}$, where $T$ is the total number of training steps, with $k_0 = d$ and $k_T = k$, and apply
\begin{equation}
\boldsymbol{A}_{\mathrm{s}}^{(t)}
=
\mathcal{S}_{k_t}(
\boldsymbol{A}^{(t)})
\end{equation}
after predetermined training steps $t$. 
This gradual pruning lets all latent dimensions participate early, when the model is still dense, and only later encourages specialization into a high-dimensional sparse representation. 
The choice of $\{k_t\}$ schedule (Figure~\ref{fig:results}a) directly affects the attainable sparsity–accuracy trade-off.

\paragraph{\textbf{Inference with Sparse Layers}} To support efficient inference with sparse embeddings, we load both $\bar{\boldsymbol{A}}_{\text{s}}$ and
$\bar{\boldsymbol{A}}^\top_{\text{s}}$
in CSC format. Maintaining two CSC-oriented copies allows both the embedding and the de-embedding to be executed using efficient SpMV kernels~\cite{bell2008efficient}. The forward pass complexity is $\mathcal{O}(nk)$, so in addition to memory savings, sparse embeddings also yield inference speedup. 

The sparse embeddings can also be used directly as item vectors in a vector database. In this setting, only one CSR-oriented copy of $\bar{\boldsymbol{A}}_{\text{s}}$ is required, and retrieval maintains the same $\mathcal{O}(nk)$ complexity.

\setlength{\tabcolsep}{4.7pt}
\begin{table*}[]
\caption{Experimental results. 
\normalfont{We report nDCG@100 for all methods across five embedding sizes (expressed in bytes (B) for ease of comparison).
We assume single-precision (\texttt{float32}) storage for all embeddings: dense embeddings store one 4-byte value per latent factor (e.g., 256 factors~$\rightarrow$~1024\,B), while sparse embeddings store each value and its index, totaling 4+4 bytes per nonzero (e.g., 128\,nonzeros~$\rightarrow$~1024\,B; in practice, the index would use fewer bytes).
Across all datasets and compression budgets, the standard errors were at most around $0.001$.}}
\label{tab:results_ndcg}
\begin{tabular}{@{}l|crrrr|crrrr|crrrr@{}}
\toprule
 & \multicolumn{5}{c|}{\textbf{Goodbooks-10k}} & \multicolumn{5}{c|}{\textbf{MovieLens-20M}} & \multicolumn{5}{c}{\textbf{Netflix Prize}} \\
 \midrule
EASE & \multicolumn{5}{c|}{0.482, embedding size 40000 B} & \multicolumn{5}{c|}{0.425, embedding size 82880 B} & \multicolumn{5}{c}{0.394, embedding size 71076 B} \\
ELSA & \multicolumn{5}{c|}{0.489, embedding size 13000 B} & \multicolumn{5}{c|}{0.429, embedding size 3200 B} & \multicolumn{5}{c}{0.395, embedding size 9800 B} \\
\midrule
Emb. size (B) & \textbf{1024} & \multicolumn{1}{c}{\textbf{512}} & \multicolumn{1}{c}{\textbf{256}} & \multicolumn{1}{c}{\textbf{128}} & \multicolumn{1}{c|}{\textbf{64}} & \textbf{1024} & \multicolumn{1}{c}{\textbf{512}} & \multicolumn{1}{c}{\textbf{256}} & \multicolumn{1}{c}{\textbf{128}} & \multicolumn{1}{c|}{\textbf{64}} & \textbf{1024} & \multicolumn{1}{c}{\textbf{512}} & \multicolumn{1}{c}{\textbf{256}} & \multicolumn{1}{c}{\textbf{128}} & \multicolumn{1}{c}{\textbf{64}} \\ \midrule
Pruned EASE & \multicolumn{1}{r}{0.478} & 0.475 & 0.473 & 0.469 & 0.458 & \multicolumn{1}{r}{0.419} & 0.414 & 0.404 & 0.391 & 0.368 & \multicolumn{1}{r}{0.389} & 0.384 & 0.377 & 0.365 & 0.343 \\
Low-Dim. ELSA & \multicolumn{1}{r}{0.433} & 0.397 & 0.357 & 0.316 & 0.270 & \multicolumn{1}{r}{0.420} & 0.410 & 0.394 & 0.371 & 0.339 & \multicolumn{1}{r}{0.376} & 0.364 & 0.349 & 0.330 & 0.307 \\
ELSA + SAE & \multicolumn{1}{r}{0.436} & 0.428 & 0.420 & 0.413 & 0.408 & \multicolumn{1}{r}{0.391} & 0.383 & 0.378 & 0.368 & 0.353 & \multicolumn{1}{r}{0.341} & 0.331 & 0.326 & 0.320 & 0.314 \\
Compr. ELSA & \multicolumn{1}{r}{\textbf{0.491}} & \textbf{0.488} & \textbf{0.483} & \textbf{0.477} & \textbf{0.469} & \multicolumn{1}{r}{\textbf{0.427}} & \textbf{0.421} & \textbf{0.412} & \textbf{0.399} & \textbf{0.379} & \multicolumn{1}{r}{\textbf{0.393}} & \textbf{0.389} & \textbf{0.382} & \textbf{0.373} & \textbf{0.357} \\ \bottomrule
\end{tabular}
\end{table*}

\paragraph{\textbf{Interpretable Segments from Sparse Latents.}} Next, we describe how to derive interpretable item segments from the sparse latent factors. For each item $i$, we identify its \emph{dominant signed latent factor} $f(i) = (\ell^*(i), s^*(i))$, defined as
$$
\ell^*(i) = \argmax_{j \in \{1,\dots,d\}}
\bigl| \bar{\boldsymbol{A}}_{\mathrm{s}}\bigr|[i,j], \ s^*(i) = \mathrm{sign}(\bar{\boldsymbol{A}}_{\mathrm{s}}[i,\ell^*(i)]),
$$
and assign item $i$ to the initial group
$ G_{(\ell^*,s^*)} = \{\, i  \mid f(i) = (\ell^*,s^*) \}$, i.e., items with the same dominant latent factor and sign are grouped together.
For every group $G$, we use available metadata of its member items to produce a short semantic descriptor 
$s$ \footnote{We use the gpt-4.1-nano-2025-04-14 LLM with OpenAI API to describe the groups.}, which is then embedded using a sentence-embedding model
$\phi$ to obtain an embedding vector $\mathbf{v} = \phi(s)$.
We then iteratively merge groups whose descriptors are
semantically equivalent, i.e., whenever the cosine similarity
$\cos(\mathbf{v}_{G_j}, \mathbf{v}_{G_k})$ exceeds a threshold $\tau \in [-1,1]$.
Merged groups receive a new composite descriptor, and the procedure
repeats until no further merges are possible.
This yields a final set of semantic segments 
$\mathcal{C} = \{ (G_c, s_c) \}$,
each consisting of a coherent item set $G_c$ and a semantic descriptor $s_c$.
These segments provide an interpretable view of the sparse latent
space and expose an inverted-index structure aligned with the
model’s representation.

Each segment is associated with a set of signed latent
factors $L_c = \{\, (\ell^*(i), s^*(i)) \mid i \in G_c \,\}$. We can therefore construct a sparse
segment--latent matrix $\boldsymbol{B}_{s} \in \mathbb{R}^{C \times d}$,
where row $c$ contains nonzeros (values $\pm1$)
only in the dimensions specified by $L_c$. Finally, we apply row-wise $\ell_2$-normalization to obtain $\bar{\boldsymbol{B}}_{s}$, which places the segments in the same latent space as the
item embeddings and makes them compatible with ELSA's inference step. 
This allows us to compute segment relevance scores $\hat{\mathbf{r}}_{\mathrm{seg}}$ from a user's interaction vector $\mathbf{x}$ as
$$
\hat{\mathbf{r}}_{\mathrm{seg}}^\top = \mathbf{x}^{\top}\,\bar{\boldsymbol{A}}_s\,\bar{\boldsymbol{B}}_s^{\top},
$$
enabling unified
recommendation of both items and semantic item segments using a single
latent user representation.

\section{Experiments}

\noindent \textbf{\textit{Experimental Setup}.} 
We evaluated our model on three widely used recommendation datasets: \textit{Goodbooks-10k} \cite{goodbooks2017}, \textit{MovieLens-20M} \cite{movieLens}, and \textit{Netflix Prize} \cite{Bennett2007}, following the evaluation setup of \cite{sparse_elsa}, i.e., feedback binarization and strong generalization with disjoint train, validation, and test set users. Detailed setup description and reproducibility instructions are available from the project repository.

\paragraph{\textbf{Baselines}}
The proposed \textit{Compressed ELSA} was compared against (i) dense, uncompressed models \textit{EASE}~\cite{ease} and \textit{ELSA}~\cite{sparse_elsa}, (ii) \textit{Low-Dimensional ELSA}, i.e., standard ELSA trained with fewer latent factors, and (iii) sparse compression approaches. The last category contained \textit{Pruned EASE} \cite{pruned_ease}, which prunes each row of the weight matrix to retain only the $k$ largest values in absolute magnitude and 
\textit{ELSA+SAE})~\cite{compressae}, which uses a sparse autoencoder (SAE) to project learned ELSA embeddings into a sparsely activated latent space.

\paragraph{\textbf{Accuracy vs.\ Compression}}  
Table~\ref{tab:results_ndcg} presents recommendation accuracy results w.r.t.\ nDCG@100. 
Across all datasets and compression budgets, \emph{Compressed ELSA} achieves the best accuracy--size trade-off. Our approach outperforms (i) \emph{Low-Dimensional ELSA}, which suffers substantial accuracy degradation (especially at higher compression levels); (ii) post-hoc sparse compression (\emph{ELSA+SAE}); and even (iii) \emph{Pruned EASE}, which is known to represent a strong parameter-efficient baseline~\cite{10.1145/3640457.3688179,pruned_ease}. Ultimately, \textit{Compressed ELSA} attains accuracy close to uncompressed (dense) ELSA despite being up to two orders of magnitude smaller.

\paragraph{\textbf{Pruning Schedules and Embedding Width}}
We tested several configurations of Compressed ELSA's training procedure. For pruning schedules, we used $k_t = k$ (\textit{Constant}) as a baseline, and experimented with \textit{Linear} and \textit{Exponential} decays of $k_t$ after each epoch, as well as a \textit{Step-wise} schedule, where $k_t$ was reduced after 10 epochs. The step-wise schedule resembles the approach in the Lottery-Ticket Hypothesis (LTH) paper~\cite{frankle2019lotterytickethypothesisfinding} which proposes training networks to convergence before increasing sparsity. We also compared two ways of proceeding after each pruning step:
(a) \textit{restarting} training from the original initialization of the remaining parameters (as in~\cite{frankle2019lotterytickethypothesisfinding}), or 
(b) continuing training from the pruned network.

We found that decreasing $k$ after each epoch performs better than waiting for full convergence, contrary to the LTH recommendation~\cite{frankle2019lotterytickethypothesisfinding} (see Figure~\ref{fig:results}b and extended results in our demo). 
On the other hand, restarting from the original initialization yields slightly better results than continuing from the pruned network, consistent with prior findings~\cite{frankle2019lotterytickethypothesisfinding}.

Finally, we evaluated the effect of initial embedding dimensionality $d$. Figure~\ref{fig:results}c shows results for the \textit{Exponential} pruning strategy \textit{with restarts}, varying $d \in \{2048, 4096, 6144, 8192\}$. 
We find that starting from larger embeddings generally yields higher-quality compressed representations for the same final embedding size, and in some configurations even surpasses the performance of full dense ELSA. However, we observe diminishing returns as $d$ increases, suggesting that an optimal training cost--performance setting may be close to the one selected for our initial experiments ($d = 4096$).
\begin{figure}[t]
\includegraphics[width=\linewidth]{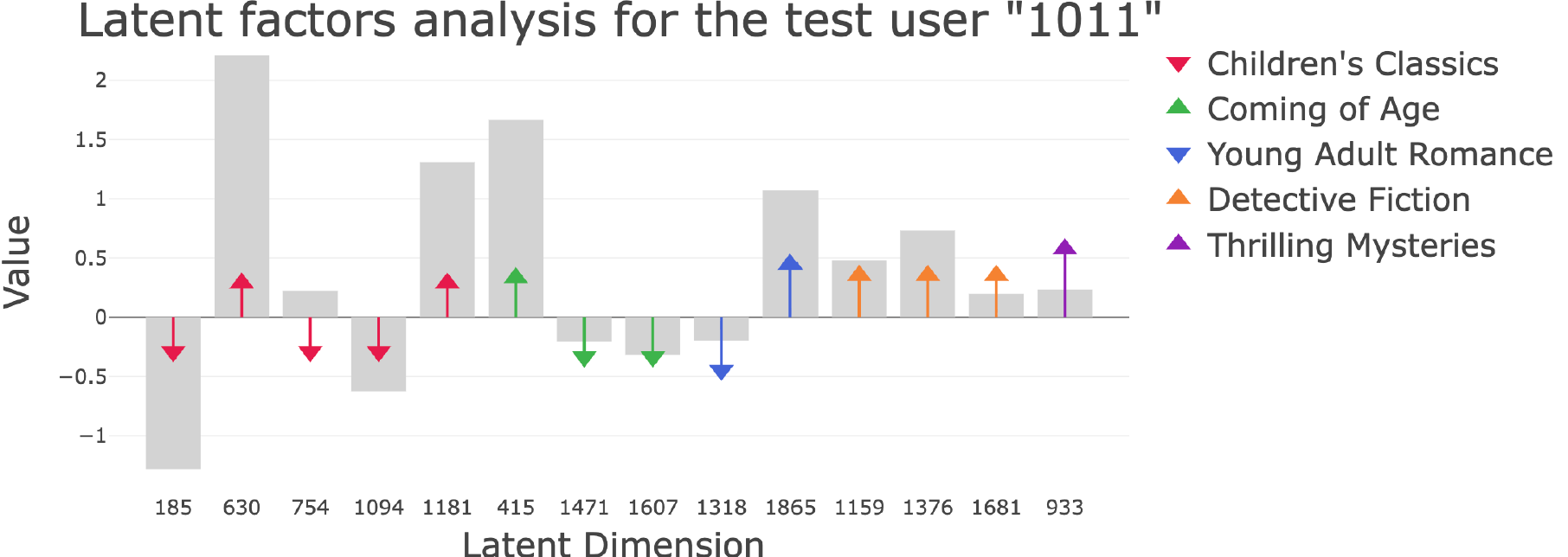}
\caption{Agreement between user activations and segment-specific latent dimensions. \normalfont{Gray bars show the user’s sparse latent-factor values, and colored arrows mark dimensions linked to semantic segments recommended to this user. The arrows consistently align with the user’s activation values, indicating a match in segment-level preferences.
\emph{(Figure extracted from our online demo.)}} }
\label{fig:latents}
\end{figure}

\paragraph{\textbf{Segment Interpretability in Practice.}} To evaluate whether the sparse latent structure learned by
Compressed ELSA produces meaningful and coherent segments, we apply
the segmentation procedure described in Section~\ref{method}
to the Goodbooks-10k dataset.  
We observe that items sharing dominant latent factors with the same polarity form semantically coherent groups such as \emph{Children’s
Classics}, \emph{Detective Fiction}, or \emph{Thrilling Mysteries}. Note that these item segments emerge 
purely from user interactions, i.e., no metadata used during training.
To verify that the segments align with user preferences, we inspect
the latent activation patterns of test users. 
By construction, 
the activation vector $\mathbf{x}^{\top}\,\bar{\boldsymbol{A}}_s$
assigns positive mass to those latent dimensions associated with
the user’s interacted segments which we observe generally align well with their
\emph{recommended} segments (see Figure~\ref{fig:latents}), and items from those segments consistently appear among the user's top $N$ recommendations.
This reveals a coherent chain of explanation from latent
activations through segment relevance to item-level
recommendations.

This is further illustrated in our interactive demo (\url{http://bit.ly/4oSi5Pj}),
where one
can inspect latent activations for different users and their corresponding
segment recommendations produced by $\mathbf{x}^{\top}\,\bar{\boldsymbol{A}}_s\,\bar{\boldsymbol{B}}_s^{\top}$.  
Qualitative inspection confirms that the sparse latent space learned
by Compressed ELSA produces interpretable, semantically coherent
segments aligned with real user behavior.

\section{Conclusions and Limitations}
We introduced Compressed ELSA, a sparse and scalable variant of ELSA that matches the accuracy of the original model while achieving orders-of-magnitude compression. Our results show that post-hoc sparsification (e.g., CompresSAE) degrades quality, whereas gradual pruning schedules preserve accuracy even at high compression levels. The resulting sparse activations also uncover coherent, interpretable item segments aligned with user behavior.

A limitation of our approach is that segment descriptors rely on an
external semantic model 
for labeling latent factors, introducing a dependency on a language model (even though this step is lightweight).
In addition, our interpretability evaluation is qualitative and centered on Goodbooks-10k; assessing segment coherence at larger scale is an important direction for future work.

Future work will explore whether the proposed gradual
sparsification strategies generalize to non-linear
architectures such as transformer-based models, whose greater sensitivity to gradient updates may necessitate further methodological refinement.

\begin{acks}
This paper has been supported by the Czech Science Foundation (GA\v{C}R) project 25-16785S.
\end{acks}

\bibliographystyle{ACM-Reference-Format}
\bibliography{sample-base}

\end{document}